\documentclass[prl,final,twocolumn,showpacs]{revtex4}

\usepackage{graphicx}
\usepackage{dcolumn}

\begin{document}

\title{Different wavelength oscillations in the conductance of $5d$
  metal atomic chains}

\author{L. de la Vega$^1$} \author{A. Mart\'{\i}n-Rodero$^1$} \author{A. Levy
  Yeyati$^1$} \author{A. Sa\'ul$^2$}

\affiliation{$^1$Departamento de F\'{\i}sica Te\'orica de la 
  Materia Condensada C-V. \\
  Universidad Aut\'onoma de Madrid. E-28049 Madrid. Spain.}

\affiliation{$^2$Centre de Recherche sur les M\'ecanismes de la
  Croissance Cristalline, CNRS, \\Campus de Luminy, Case 913, 13288
  Marseille Cedex 9, FRANCE}

\date{\today}

\begin{abstract}
  Combining ab-initio and self-consistent parametrized tight-binding
  calculations we analyze the conductance properties of atomic chains
  of $5d$ elements like Au, Pt and Ir.
  It is shown that, in addition to the even-odd parity oscillations
  characteristic of Au, conduction channels associated with the almost full
  $d$ bands in Pt and Ir give rise to longer periods which could be
  observed in sufficiently long chains. The results for short chains
  are in good agreement with recent experimental measurements.
\end{abstract}


\maketitle
   
A traditionally idealized textbook example, the atomic linear chain,
has recently become an actual system that can be explored
experimentally.  The formation of atomic chains several (up to 7-8) atoms 
long has been achieved in recent years using experimental techniques like
scanning tunneling microscope and mechanically controllable break
junctions \cite{review}. In both cases, evidence has been found of
chain formation in the last stages of pulling an atomic contact for
certain metallic elements like Au, Pt and Ir.  It has been suggested
that the stability of the chains for these $5d$ elements as opposed to
other metals is related to relativistic effects involving transfer of
charge between $sp$ and $d$ bands \cite{smit01}.

These atomic chains exhibit, on the other hand, interesting
conductance properties.  Recent experimental results indicate the
presence of conductance oscillations with period $\sim 2a$ ($a$ being
the interatomic distance) after averaging over many realizations of
the chain \cite{smit03}.  While in the case of Au the oscillations are
superimposed to an almost constant background of the order of the
quantum of conductance $G_0 = 2e^2/h$, in the case of Pt the mean
value of the conductance exhibits a continuous decrease from $\sim
2.5$ to $\sim 1.0 G_0$.  On the other hand, for Ir chains the
conductance varies between $\sim 2.2$ and $\sim 1.8 G_0$ with a less
clear oscillatory behavior.

Theoretical studies of the conductance of atomic chains have been so
far mainly restricted to idealized models \cite{pernas} or to
ab-initio calculations for light metallic elements like Na, Si or C
\cite{Na-C-Si}, but realistic calculations for the actual $5d$ metal
chains, in which the role of $d$ orbitals is expected to be relevant,
are still lacking \cite{comment-Au}. From the theoretical point of
view even-odd conductance oscillations are already present at the
level of simple tight-binding (TB) models for the half-filled case
\cite{pernas}.  This is an interference phenomena arising from the
commensurability of the Fermi wavelength and the lattice spacing.
This behavior has also been observed in more realistic calculations
based on ab-initio methods for Na and C \cite{Na-C-Si,comment-Al}.
This simple explanation can account qualitatively for the behavior in
the case of Au, characterized by a full $5d$ band and a nearly
half-filled $6s$ band. 

For single atom Pt contacts the value around 2.5 $G_0$ has been
attributed to the contribution of additional channels associated with
the $5d$ orbitals \cite{Pt-oneatom}.  Thus the observation of
oscillations with period $2a$ is not at all obvious in this case and
deserves a detailed analysis.  The aim of this work is to understand
the possible mechanisms leading to this behavior. 
We will show that the channels associated with the almost full $d$ 
bands in Pt and Ir exhibit transmission oscillations with longer 
periods which can be associated with the Fermi wave vector of the
bands in the infinite chain. 

\begin{figure}
\includegraphics[width = 6.5 cm]{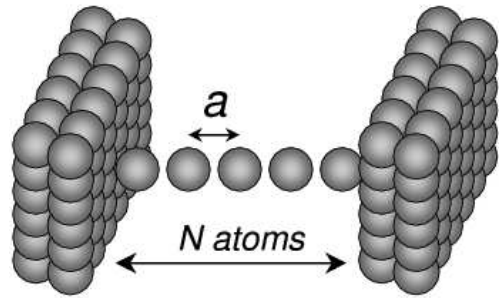}
\caption{Model geometry considered for the calculation of the
conductance. The electrodes are represented by semi-infinite fcc 
crystals grown in the (111) direction.}
\label{fig:geom}
\end{figure}

We shall consider model geometries like the one depicted in Fig.\ 
\ref{fig:geom} in which the atomic chain is connected to bulk
electrodes represented by two semi-infinite fcc perfect crystals grown
along the (111) direction.  The interatomic distance $a$ will be
allowed to take larger values than in the relaxed structure of the
infinite chain to account for the effect of the applied stress. 
For simplicity we shall concentrate first in the analysis of the ideal linear
geometry discussing in a second step the effect of 
possible structural deformations.

A convenient approach to quantum transport calculations in atomic size
conductors uses a self-consistent parametrized TB model in which the
main features of the bulk band structure are included \cite{us98}.
This type of calculation has been fruitful to obtain information
about the conduction channels in one atom
contacts \cite{nature}. 
In the particular case of atomic chains a natural starting
point is the band structure of the infinite linear chain whose main
properties are discussed below.

The electronic structure and total energy calculations for the 
infinite chains
were performed using the {\sc WIEN97} code \cite{wien97}. This is an
implementation of the linearized augmented plane wave method based on
density functional theory. For exchange and correlation we have used
the local density approximation (LDA), with the correlation part as
given by Perdew and Wang \cite{lsda5}. 

\begin{figure}
\includegraphics[width = 7 cm]{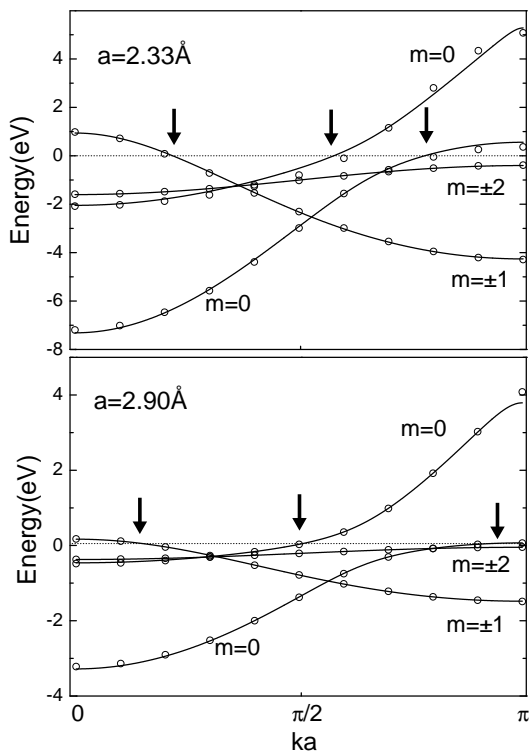}
\caption{Band structure of the infinite Pt chain for two different interatomic
  distances. The symbols and the full lines correspond to the
  ab-initio and to the fitted TB bands respectively.  The bands are
  classified by the quantum number $m$ corresponding to the projection
  of the angular momentum on the chain axis. The arrows indicate the
  crossing of the Fermi level for the $m=0$ and the $m=\pm 1$ bands.}
\label{fig:band-struct-pt}
\end{figure}

We will discuss the band structure of the chains
taking the case of Pt as a representative example (Au and Ir exhibit a
rather similar band structure although displaced with respect to
$E_F$).  Fig.\ \ref{fig:band-struct-pt} shows the bands around the
Fermi energy for Pt at two different situations with increasing
stress: one close to the calculated equilibrium situation 
($a=2.33$ \AA) and the other 
corresponding approximately to the maximum stress before breaking the
chain ($a=2.90$ \AA).
The fits of these bands using a Slater-Koster TB scheme are
shown as full lines.  Several features of this band structure are
worth commenting: 
1) Symmetry considerations allow to classify the bands according to the
projection of the angular momentum along the chain axis $m$ (we take
the $z$ axis along the chain direction).
2) Close to the Fermi level there is an almost flat filled two-fold
degenerate band with $d_{xy}$ and $d_{x\sp2-y\sp2}$ ($m=\pm 2$)
character.  The other partially filled and more dispersing bands have
$s-p_z-d_{z\sp2}$ ($m=0$) and $p_x-d_{xz}$ or $p_y-d_{yz}$ ($m=\pm 1$)
character (see labels in Fig.\ \ref{fig:band-struct-pt}).
3) When the chain is elongated the dispersion of the bands is reduced
and a net charge transfer from $s$ to $d$ is observed, tending rather
fast to the ``atomic" charge configuration $5d^9 6s^1$.  As a consequence the
upper $m=0$ band with a predominant $s$ character gets closer to the
half-filled situation ($k_Fa \sim \pi/2$) which, as commented above, 
would give rise to even-odd
conductance oscillations. Notice on the other hand
that the $m=\pm1$ bands cross the Fermi level at a lower wave vector. 

In order to describe the electronic properties for a model geometry
like the one of Fig. \ref{fig:geom} the TB Hamiltonian is written as
$\hat{H} = \sum h_{i,j,\alpha,\beta}
\hat{c}^{\dagger}_{i,\alpha,\sigma} \hat{c}_{j,\beta,\sigma}$ $=$
$\hat{H}_{chain}$ $+$ $\hat{V}_L$ + $\hat{V}_R$ + $\hat{H}_L$ +
$\hat{H}_R$, where $\hat{H}_{chain}$ and $\hat{H}_{L,R}$ describe the
uncoupled chain and the left and right electrodes respectively;
$\hat{V}_{L,R}$ being a coupling term between the chain and the
electrodes which, in the model geometry of Fig. \ref{fig:geom}
corresponds to the hopping elements connecting the outermost atoms of
the chain and their three nearest neighbors on the electrodes surface.
The matrix elements $h_{i,j,\alpha,\beta}$, where $i,j$ and
$\alpha,\beta$ design sites and orbitals respectively, are taken from
the fits to the chain and to the bulk {\it ab-initio} band structure.  As a
self-consistency condition we impose local charge neutrality by fixing
the charge of the $d$ orbitals and $sp$ orbitals within the chain
equal to the ones of the infinite chain \cite{orb-charge-neu}. 
The corresponding condition is also imposed on the electrodes surfaces.  

Once the TB Hamiltonian has been built the conductance $G(E)$ = $G_0
\mbox{Tr} \left[\hat{t}^{\dagger}(E) \hat{t}(E) \right]$ of the chain
is calculated in terms of the matrix elements of the Green function
operator $\hat{G}^{r}(E)$ =
$\lim_{\eta\rightarrow 0} \left[E + i\eta - \hat{H}\right]^{-1}$ using
:
\begin{eqnarray}
\hat{t}(E) = 2 \hat{\Gamma}^{1/2}_L(E) \hat{G}^{r}_{1N}(E) 
\hat{\Gamma}^{1/2}_R(E),
\end{eqnarray}
where $\hat{\Gamma}_{L,R}$ are the matrix tunneling rates connecting
the chain to the leads \cite{us98}.  The transmission matrix
$\hat{t}^{\dagger}\hat{t}$ can then be diagonalized in order to obtain
the transmission eigenchannels and eigenvalues $\tau_n$ which
characterize completely the linear transport properties of the system.

\begin{figure}
\includegraphics[width = 8 cm]{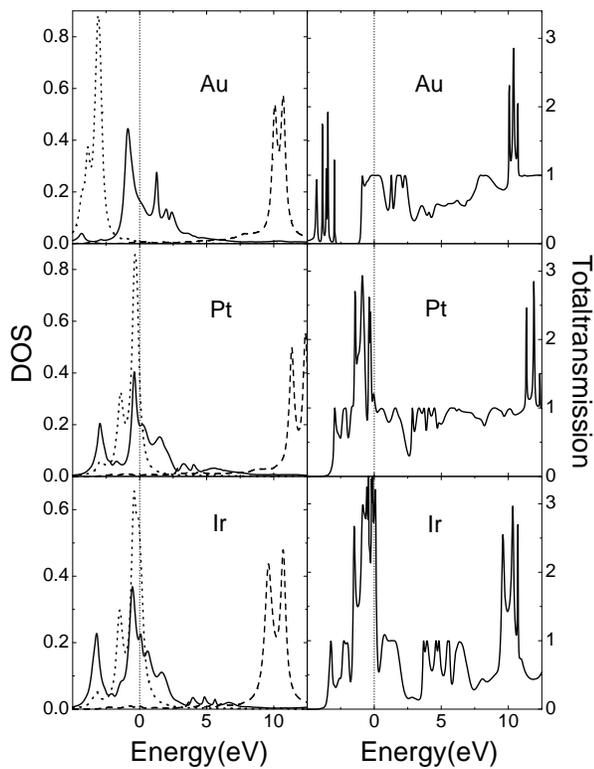}
\caption{Local density of states (LDOS) at the central atom 
  and total transmission for Au, Pt and Ir chains with $N=5$ at an
  intermediate elongation.
  The LDOS is decomposed in $s$ (full line) $d$ (dotted line)
  and $p$ dashed line with the same normalization in the three cases.}
\label{fig:dos-and-transm}
\end{figure}

In order to illustrate the main features and the differences between
Au, Pt and Ir, we show in Fig.\ \ref{fig:dos-and-transm} the density
of states and the energy dependent transmissions for a $N=5$ chain of
these metals at an intermediate elongation. 
As has been shown in previous works \cite{comment-Au},
Au chains are characterized by a single conducting channel around the
Fermi energy with predominant $s$ character. The transmission of this
channel lies close to one and exhibits small oscillations as a
function of energy resembling the behavior in a single band TB model
\cite{pernas}. For odd $N$ charge neutrality provides a strong
mechanism for the almost perfect quantization of the conductance, as
discussed in Ref. \cite{us97} for one atom. 

In the case of Pt the contribution from the almost filled $5d$ bands
becomes important for the electronic properties at the Fermi energy.
There are three conduction channels with significant transmission at
$E_F$: one due to the hibridization of $s-p_z$ and $d_{z\sp2}$
orbitals, and other two almost degenerate with $p_x-d_{xz}$ and
$p_y-d_{yz}$ character respectively.  One can naturally associate this
channel distribution with the band structure presented in Fig.\ 
\ref{fig:band-struct-pt}.

The contribution of the $5d$ orbitals is even more important in the
case of Ir where a fourth channel related to the lower $m=0$ band
exhibits a significant transmission.

\begin{figure}
\includegraphics[width = 7 cm]{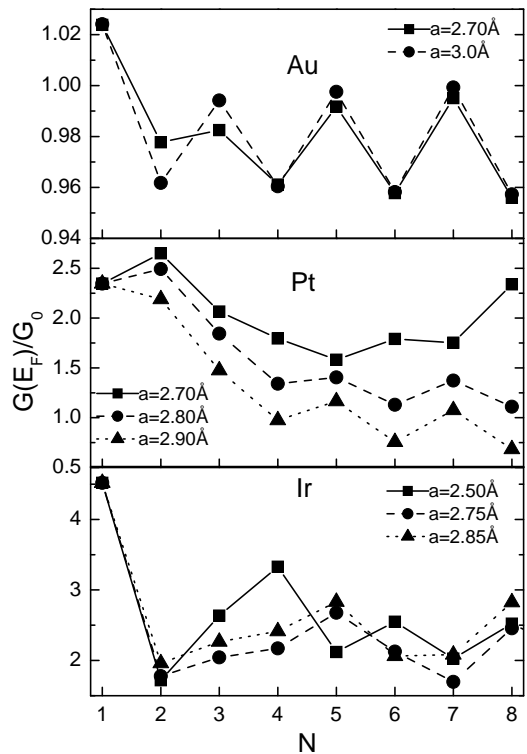}
\caption{Evolution of the conductance with $N$ for different values
  of the interatomic distance $a$.}
\label{fig:gvsn}
\end{figure}

Let us analyze the evolution of the conductance with the number of
atoms in the chain. We find that this evolution is rather sensitive to
the elongation, specially in the case of Pt and Ir (for Au
the conductance exhibits small amplitude even-odd oscillations, $\sim
0.04 G_0$, which remain basically unaffected upon stretching).  The
center panel in Fig.\ \ref{fig:gvsn} shows the total conductance of Pt
chains as a function of $N$ for different interatomic distances.  Notice
that a slight increase of $a$ from the equilibrium value leads to a
global decrease of the conductance with $N$ together with the
appearance of even-odd oscillations. For $a=2.9$ \AA\ the total
conductance drops from $2.4 G_0$ to around $1 G_0$ for $N >4$, while
the oscillations amplitude is of the order of $0.2 G_0$.

\begin{figure}[t]
\includegraphics[height = 9.5 cm]{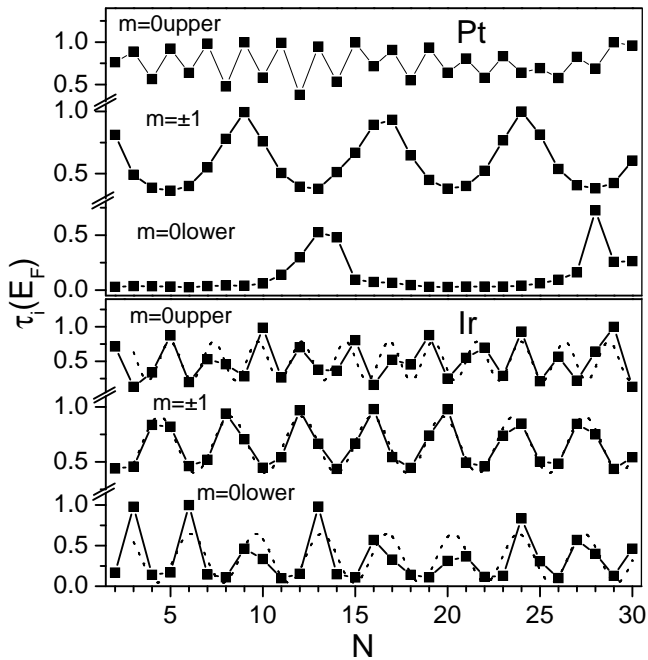}
\caption{Channel decomposition of the conductance for Pt and Ir 
  as a function $N$ for intermediate elongations. The legends indicate the
  symmetry of the associated bands in the infinite chain. The dotted
  lines in the case of Ir correspond to functions of the form $\sim
  \cos^2(k_{F,i}Na)$ where $k_{F,i}$ is the Fermi wave vector of the
  associated bands, which help to visualize the oscillatory behavior.}
\label{fig:chan-decom-pt}
\end{figure}

A deeper understanding of this behavior can be gathered when analyzing
the evolution of the conductance and its channel decomposition for
even longer chains. 
This is illustrated in Fig.\ \ref{fig:chan-decom-pt} for the case of 
Pt and Ir. As can be
observed the decrease of the total conductance of Pt for $N < 7-8$
corresponds actually to a long period oscillation in the transmission of
the two nearly degenerate channels associated with the $m = \pm 1$
bands. This period can be related to the small Fermi wave vector 
of these almost filled $d$ bands, as indicated by the left 
arrows in Fig.\ \ref{fig:band-struct-pt}. 
In addition, the upper $m=0$ band crossing the Fermi level is close to
half-filling giving rise to the even-odd
oscillatory behavior observed in the transmission of the channel with
predominant $s$ character.
The lower $m=0$ band tends to be completely filled as
the chain is elongated and the corresponding channel is nearly closed 
for short chains.
However, for intermediate elongations one can still appreciate a very
long period oscillation in its transmission, rising up to $\sim 0.5G_0$, 
as can be observed in the upper panel of Fig.\ \ref{fig:chan-decom-pt}.

In the case of Ir both the $m=\pm 1$ and the lower $m=0$
are less filled than in Pt leading to four partially open channels.
The corresponding values of the Fermi wave vector fix its overall
oscillatory behavior as illustrated in the lower panel of 
Fig.\ \ref{fig:chan-decom-pt}. 

The results presented so far correspond to the idealized geometry
depicted in Fig.\ \ref{fig:geom}. Deviations from this situation can
arise both from disorder in the atomic positions and possible 
structural distortions. Previous studies indicate that the relaxed 
structure for Au chains corresponds to a zig-zag configuration 
\cite{sanchez-portal}.
Our ab-initio calculations indicate that this is
also the case for Pt and Ir, the main effect of the external
stress being first to suppress the zig-zag deformation and then to
increase the interatomic distances. 
This deformation has some effect on the band structure, mainly 
consisting in the breaking of the degeneracies present in the linear case.   
Although this would affect the results close to the relaxed situation,
we have checked that the main conclusions regarding the 
appearance of long period oscillations associated with the $d$ bands
are still valid. On the other hand, molecular dynamics simulations
suggest a non-uniform distribution of bond lengths along the chain
\cite{dasilva}. We have found that this type of deformation as well as
the inclusion of moderate disorder in the atomic positions do not
give rise to qualitative changes in the conductance behavior as long
as they remain within the ranges predicted in Ref.
\cite{dasilva}.  

In conclusion we have analyzed the conductance properties of Au, Pt
and Ir atomic chains. In addition to the well understood even-odd
parity effects of Au, in metals like Pt and Ir 
the partially filled $5d$ bands give rise to a more complex oscillatory
pattern in which more than a single wavelength can be identified. 
As a general rule we predict that the transmission corresponding to each
conduction channel oscillates as $\sim \cos^2(k_{F,i}Na)$ where 
$k_{F,i}$ is the Fermi wave vector of the associated band in the
infinite chain. In the case of Pt the charge transfer from $s$ to
$d$ orbitals that takes place upon stretching leads to a nearly
half-filled $s$ band with the consequent even-odd oscillations 
in the corresponding channel, while the channels associated with the
almost filled $d$ bands exhibit longer period oscillations. 
The total conductance for short chains ($N < 7-8$) 
exhibits an overall decrease with
superimposed even-odd oscillations in qualitative agreement
with the experimental results of Ref.  \cite{smit03}.   
In the case of Ir we find four partially open channels with rather
similar wavelengths leading to a less clear oscillatory pattern. 
Our predictions for the channel decomposition 
could be checked experimentally by similar techniques as the ones
used in Refs. \cite{nature,gabino}.
Advances in the fabrication techniques would allow to test
these predictions for longer chains.

This work has been partially supported by grant BFM2001-0150 (MCyT, Spain).  
The authors would like to thank J.M. van Ruitembeek for fruitful
comments.

\end{document}